# Evaluating an assembly- and disassembly-oriented expansion of Modular Function Deployment through a workshop-based assessment

Fabio Marco Monetti[a]*

[a]*KTH Royal Institute of Technology, Brinellvägen 8, 100 44, Stockholm, Sweden*

* Corresponding author. Tel.: +46 73 461 8925. *E-mail address:* monetti@kth.se.

**Abstract**

Modular product architectures are used to enhance flexibility, reduce production complexity, and support sustainability goals. However, traditional Modular Function Deployment (MFD) method does not fully integrate Design for Assembly (DFA) and Design for Disassembly (DFD) principles, leading to sub-optimal manufacturability and end-of-life strategies. This study introduces an expanded MFD method incorporating assembly and disassembly considerations into early-stage modularisation. A workshop-based evaluation assesses usability and applicability, involving participants using standard and expanded MFD. Results indicate that integrating DFA and DFD enhances assembly efficiency, ease of disassembly, and modular product strategy alignment. However, usability challenges were identified, necessitating refinements for industry application.

## 1. Introduction

Modularisation is a strategic design approach that divides products into logical building blocks, enabling production scalability and flexible product configurations to enhance customer value [1]. It reduces internal complexity, expands external variety, and supports faster lead times, lower costs, and easier redesign [2]. It also benefits sustainability and maintenance by enabling repair, upgrade, and recycling, thus reducing waste and promoting long-term module reuse [3].

Creating modules alone is insufficient to realise modularity's full potential. Their organisation must reflect lifecycle needs, as assembly and replacement strategies directly impact customer and company benefits [4]. Realising this potential requires streamlined assembly and disassembly processes, standardised interfaces, fewer components, and early definition of operations and sequences.

Central to these efforts are Design for Assembly (DFA), reducing assembly time and cost while improving quality [5,6] and Design for Disassembly (DFD), which facilitates efficient end-of-life repair, reuse, and recycling. The latter focuses on non-destructive disassembly, easy separation, and recoverability [7], and both are most impactful when applied early, during concept development.

Despite their value, modularity integrating DFA and DFD is complex, and requires coordination across departments and significant know-how [2]. These demands highlight the need for effective tools and methods to support decision-making [8].

Modular Function Deployment (MFD) [9] was originally developed to align modular design with strategic targets, but still lacks full integration of DFA and DFD. This study proposes an expanded MFD approach that embeds assembly and disassembly considerations into early product architecture development, aiming to reduce cost, improve assembly, and support maintenance in complex systems.

While modular architecture must ultimately support goals like usability, maintainability, and sustainability, this study prioritises assembly and disassembly for their direct influence on production and end-of-life. Many of the resulting decisions contribute indirectly to broader lifecycle benefits.

This paper presents and evaluates an expanded MFD method. The assessment focuses on usability and applicability to understand its practical value. This aligns with [10], who emphasise iterative evaluation in tool design. Results provide feedback to refine the addition of DFA and DFD in modularity.

## 2. Literature review

In traditional product design, customer requirements are translated into product features with defined cost targets. This leads to complex assemblies of dependent parts, optimised for cost or performance [11]. Keeping complexity is resource-intensive, requiring frequent updates to accommodate market changes, new technologies, and component obsolescence [12].

Modular product design addresses these challenges by aligning development with strategic goals. It enhances flexibility, reduces lead times and costs, and supports production efficiency and product variety [1,13]. Achieving optimal architectures, however, requires synced input from strategic planning, market insight, and engineering teams [8].

Adding production-oriented criteria early can simplify products, reduce assembly time, improve quality, and accelerate time-to-market [6]. DFD supports sustainability goals by enabling recyclability and end-of-life separation [14]. Together, DFA and DFD are vital for aligning product architecture with manufacturing and maintenance objectives.

Despite this, incorporating DFA and DFD during early design stages remains difficult due to the qualitative nature of early data [15]. Their overlap allows for shared input, yet their different implications (differing connector, or joint strategies) require distinct treatment [16]. Most existing methods separate these, missing chances for integrated decision-making [17].

MFD is a widely used modularisation method, especially in Sweden. It follows structured steps: capturing customer needs, identifying product functions, and applying module drivers [9]. While refined over time [18,19], it does not natively integrate DFA or DFD, often resulting in architectures that overlook assembly or disassembly efficiency. DFA is typically introduced late in design, incurring higher costs [20], and organisational misalignment further limits integration [8]. Other modularisation methods, as DSM [21], offer structuring logics but lack explicit integration of lifecycle-oriented criteria. The expanded method addresses this gap by embedding DFA and DFD considerations during concept development.

This study proposes a few support tools that extend MFD to embed DFA and DFD considerations into modular architecture development. The goal is to evaluate its usability and applicability in a controlled workshop setting reflecting realistic product development conditions. Hence, the following research question (RQ) is posed: *how effective, in terms of usability and applicability, is a new support tool that includes the integration of Design for Assembly and Design for Disassembly principles within the Modular Function Deployment method?*

## 3. Research methodology

This study is part of a larger research programme following the Design Research Methodology (DRM) framework [10]. Previous work covered the Research Clarification and Descriptive Study I stages [8,20]; this paper is the first iteration of the Prescriptive Study. The goal is not to develop a market-ready tool but to evaluate proposed conceptual support for integrating DFA and DFD within modular product design.

The study was conducted in collaboration with a Stockholm-based consultancy specialised in modularity, which applies MFD in industrial settings and teaches the MG2020 modularisation course at KTH Royal Institute of Technology.

The workshop approach was chosen to evaluate usability and applicability through hands-on interaction, immediate feedback, and iterative refinement in a realistic design setting. This aligns with [22], who highlight the value of immersive formats for exploring new design supports.

Established workshop guidelines were followed [23], with a focus on assessing how well the expanded MFD tool supported participants in navigating DFA and DFD trade-offs. The author provided instructional guidance and led the qualitative analysis, while facilitators documented field observations throughout the process.

Triangulation was achieved using multiple data sources. *Pre-* and *post-workshop questionnaires* captured changes in participants' understanding of modularisation, DFA, and DFD. *Field observations and notes* recorded observations on engagement, tool use, and team dynamics. *Group brainstorming outputs* recorded observations on engagement, tool use, and team dynamics. *Participant artifacts* included completed matrices and design sketches. *Time tracking* assessed task duration and workflow fluency.

*Usability* was defined as the ease of understanding and applying the tool; *applicability* referred to its relevance for solving modularisation challenges. Usability was analysed via facilitator involvement, observed uncertainty, and timing. Applicability was measured through artefact depth, feedback, and produced ideas. A six-step thematic analysis followed [24].

The workshop was conducted over two consecutive three-hour sessions at KTH's Industrial Production Department. Participants were Master's students enrolled in MG2020, randomly assigned to one of three pairs. All had equivalent theoretical training in MFD. A pilot session with different participants was run beforehand to refine the workshop format.

The design task involved proposing an improved modular concept for the leaf blower, with the goal of enhancing either assembly or disassembly performance based on the assigned criteria. Design freedom was limited to module structure, interfaces, and connectors, rather than full functional redesign.

A technician with extensive experience in assembly and disassembly processes, facilitated the initial hands-on disassembly of a manually operated leaf blower (see Fig. 1), allowing all participants to assemble and disassemble the product individually. This was followed by a group brainstorming session to identify functional challenges and improvement areas.

Participants were then split into three specific groups.
- Group A utilised the standard MFD support tool.

- Group B applied an expanded MFD tool with DFA-oriented evaluation criteria (e.g., reduced part count, insertion ease, minimal reorientation and tooling [6]).
- Group C used DFD-oriented criteria (e.g., easy removability, connector standardisation, accessibility, damage avoidance [14]).

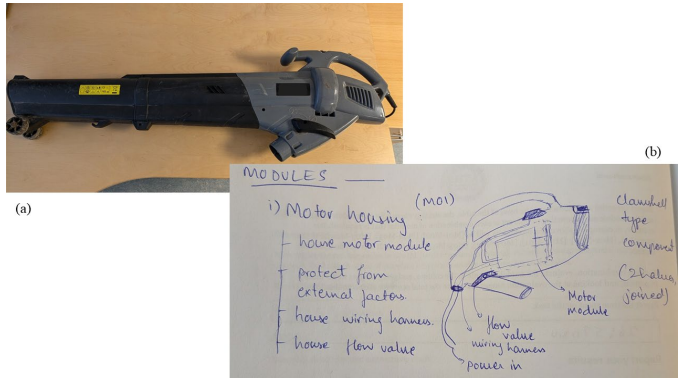

Fig. 1. Workshop context: (a) the leaf blower used as the case product; (b) participant-generated sketch exploring module logic and connection requirements. This exemplifying artefact illustrates the hands-on nature of the task and translation of disassembly/assembly insights into modular concepts.

Each group completed the full expanded MFD method using printed templates and digital worksheets. Participants received an introductory briefing with slides on tool and matrix logic. Groups B and C received short additional instructions and printed scoring sheets specific to DFA or DFD, respectively.

All groups followed the same MFD steps, but the evaluation criteria for concept evaluation and MSASM varied. Group A defined their own criteria; Groups B and C received formalised guides. These influenced both structured scoring and qualitative reasoning.

Each group worked independently. Pairs collaborated internally but no information was exchanged across teams. This segmentation reduced cognitive load and allowed focused exploration of assembly vs disassembly. Field notes captured tool usability and group dynamics. All analysis was conducted using thematic coding to identify key challenges, successes, and usage patterns.

All participants provided informed consent and were assured of confidentiality and data protection, adhering to standard research ethics protocols.

## 4. Modular Function Deployment (MFD), expanded

The original MFD method [9], defines modular architectures through five key steps: (i) clarify requirements, (ii) select technical solutions, (iii) generate concepts, (iv) evaluate them, and (v) refine modules. The process is iterative, and the Product Management Map (PMM) [25] structures data into supporting matrices.

This paper presents an expanded, action-oriented MFD version. It retains the core steps but adds tools and criteria to embed DFA and DFD principles. Fig. 2 outlines the modified PMM and highlights new additions.

Specifically, the extension includes: (i) predefined internal criteria in concept evaluation; (2) the Assembly Directions and Connections Draft (ADCD); and (iii) the Module Set Assembly Strategy Matrix (MSASM). These additions aim to improve early insight in manufacturability and end-of-life performance.

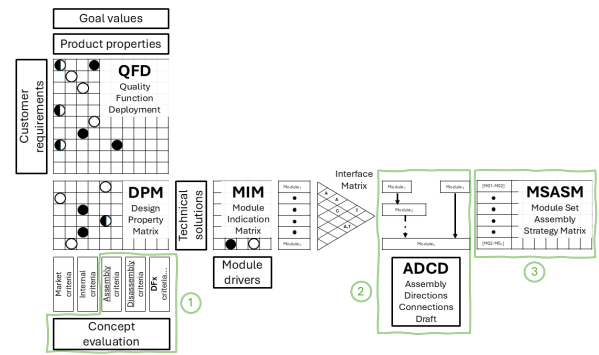

Fig. 2. Overview of the expanded MFD workflow. The original PMM is extended to: (1) production-oriented concept evaluation criteria; (2) assembly directions and connections draft; (3) module set assembly strategy matrix.

While DFA/DFD criteria were provided to the focused teams, their impact lies in how the tools structured design reasoning. Rather than informal guidance, the workflow required systematic trade-offs across competing production constraints, encouraging deeper architectural reflection.

### 4.1. Production-oriented concept evaluation criteria

In traditional MFD, concepts are evaluated based on customer needs or designer experience. The expanded method introduces structured criteria drawn from DFA and DFD principles [5,26], ensuring manufacturing and disassembly implications are considered early. Common DFA criteria include part count, tool needs, symmetry, and insertion ease; DFD includes disassembly time, connector type, and damage risk. A Pugh matrix [27] or numeric scoring can be used. Table 1 lists some selected criteria used in the workshop.

### 4.2. Assembly Directions Connections Draft (ADCD)

The ADCD tool provides a sketch-based visualisation of assembly directions, connection types, and module layout. After modules and interfaces are defined, users annotate directionality, fastener types, and potential issues like tool access or simultaneous insertion.

This tool addresses two key challenges: (i) assembly direction optimisation (to minimise reorientation and improve automation compatibility); and (ii) connection strategy (snap-fit, screws, adhesive). For example, snap-fits aid fast assembly, while screws offer durability.

The ADCD identifies inefficiencies and informs architectural refinements to improve ease of assembly or disassembly.

### 4.3. Module Set Assembly Strategy Matrix (MSASM)

The MSASM enables evaluation of module pair readiness for assembly, based on scores across production-relevant criteria. Each module set (e.g., M02–M01) is scored 1–5 per criterion (1 = optimal, 5 = poor). Scores are based on group consensus during handling and discussion. If uncertain, conservative values are used. Final scores reveal interface bottlenecks and guide redesign. Colour-coded matrices helped visualise complex areas.

For example, Group A rated the housing–motor interface poorly due to accessibility (3), orientation (2), and fastening complexity (3), flagging it as acceptable, but to revise. Table 2 summarises three core MSASM criteria spanning cognitive, technical, and ergonomic challenges.

Table 1. Selection of concept evaluation criteria used during the workshop to represent core DFA and DFD principles. The six here were chosen to highlight representative dimensions across both assembly and disassembly priorities, including procedural time, tool use, manual effort, and physical access. Each criterion was introduced in the matrix-based evaluations to support structured trade-off analysis during modular concept development.

| Criterion | Description | Type |
|---|---|---|
| Assembly time | Total time to complete assembly | DFA |
| Ease of insertion | Ease of inserting the part into its location | DFA |
| Tool requirements | Tools needed (e.g., torque tools, magnifiers) | DFA/DFD |
| Access | Accessibility of insertion/detachment point | DFA/DFD |
| Connector destruction | Must connectors be destroyed during disassembly? | DFD |
| Force intensity | Force required to detach component | DFD |

Table 2. Representative evaluation criteria used in MSASM. These criteria illustrate how module sets were assessed based on assembly feasibility and interface complexity. Each was scored on a 1–5 ordinal scale, where 1 represents optimal performance (e.g., simplicity, ease, accessibility) and 5 indicates significant assembly challenges. The selected examples cover some key drivers of module integration effort.

| Criterion | Score = 1 (best) | Score = 3 (mid) | Score = 5 (poor) |
|---|---|---|---|
| Attachment interface connections | Few, simple connections; snap-fits like | Multiple connections but manageable | Numerous or tangled connections requires tools or cable routing |
| Assembly direction | Vertical insertion with gravity assistance | Mixed directions need reorienting module | Many directions; require turning or complex manipulation |
| Accessibility | All parts visible and reachable from standard workstation | Partial obstruction; reaching inside housing | Inaccessible without disassembly or special tools |

## 5. Results

The results are structured around the two core evaluation criteria: usability and applicability of the expanded MFD tool. Drawing from multiple data sources—including workshop artefacts, facilitator notes, and participant reflections—this section highlights how the tool supported design reasoning and decision-making in assembly- and disassembly-focused contexts.

### 5.1. Usability and applicability across data sources

The evaluation of the expanded MFD method was based on triangulated insights from pre- and post-workshop surveys, field observations, brainstorming discussions, and technical artefacts. This integration revealed how participants perceived the tool's usability and applicability in addressing DFA and DFD concerns within a modularisation context.

Participants began the workshop with limited hands-on experience in structured modularisation methods. Pre-workshop surveys indicated a basic awareness of modular design's role in improving assembly efficiency, but little confidence in applying DFA or DFD principles. Field observations confirmed this, especially for Group A, whose members encountered significant difficulties navigating the Customer Value Rating (CVR) and QFD matrices. Facilitators noted repeated clarifications and a need to restart those steps due to misinterpretation of tool logic.

Despite initial difficulties, post-workshop feedback reflected a shift in participants' confidence and clarity. Group B appreciated the structured nature of the DFA-oriented criteria and noted that they "reduced complexity and accelerated value creation." Group C reported that the scoring criteria and visual matrices helped them think more explicitly about connector types, force intensity, and safe detachment — aligning with DFD priorities. One participant remarked: "It was quite interesting and practical, which actually helped me learn more than in class lectures."

Facilitators observed increasing fluency across all groups. While Group A required sustained support, Groups B and C adapted quickly. The latter groups frequently referenced production-oriented principles, even without facilitator prompts, and were able to articulate the impact of technical decisions on assembly or disassembly quality. Challenges were noted, particularly in differentiating between technical solutions and product properties, and in scoring. Suggestions for improvement included clearer instructions, pivot-style worksheet linking, and CAD/PLM integration.

Brainstorming discussions and sketch annotations further confirmed applicability. Participants identified module misalignment, connector overload, and access issues as central barriers. They proposed standardised connectors, simplified housing, and ergonomic improvements. Technical artefacts from Group B featured modular merging and cleaner interface logic aligned with DFA, while Group C included detailed notes on safety and fastener ergonomics, supporting DFD-aligned thinking. These outputs show that participants could translate experiential insights into tangible modular concepts.

Together, these findings indicate that the tool promoted structured design reasoning while maintaining flexibility to accommodate distinct objectives. Usability challenges were present, especially at the outset, but diminished with experience. Applicability was consistently supported through qualitative outcomes and design decisions embedded in real production logic.

## 5.2. Tool performance and strategy outcomes

Differences in tool use and strategy formation emerged across groups, as captured through qualitative analysis and time-on-task tracking. The ADCD and MSASM tools in particular played key roles in shaping modular decisions and uncovering operational bottlenecks.

Group A produced limited annotations in their ADCD sketches, with minimal reference to spatial or directional constraints. In contrast, Group B incorporated insertion direction, tool needs, and fastener logic, while Group C captured accessibility and safety concerns. These sketches provided an effective visual link between module layout and production logic. In MSASM matrices, Group A's scores ranged widely (13–30), with the scores for the housing–motor set (30) indicating major assembly issues, flagged due to poor accessibility and fastening complexity. Group B's module pairs scored more favourably, reflecting reduced part counts and improved insertion logic. Group C recorded consistently lower scores (e.g., M03–M08 = 11), implying simpler disassembly, and their analysis emphasised ergonomic handling and force minimisation, scoring several module sets in the optimal range.

Participant time tracking further highlighted tool performance. Group A consistently spent more time per task, especially during CVR and QFD, aligning with facilitator notes of confusion. Group B worked efficiently, completing ADCD and MSASM quickly and with detailed output, though their omission of DPM scoring suggests a possible shortcut. Group C displayed task variability, spending longer on Concept Evaluation and MIM, consistent with their thorough approach to module planning and DFD alignment.

Together, these patterns confirm that the expanded tool not only structured early modularisation decisions but also influenced how participants prioritised and evaluated solutions. While the learning curve was real, the structure encouraged focused, goal-aligned reasoning from early stages through to module-set analysis.

## 6. Discussion

The expanded MFD method shows promise in supporting structured, production-aware design decisions by integrating DFA and DFD considerations into early-stage modularisation. Results indicate the tool was usable by participants with limited prior experience and applicable to realistic modularisation challenges.

### 6.1. Usability and applicability of the expanded MFD tool

Usability was assessed through facilitator notes, time-on-task data, and participant feedback. Group A required more support, during the CVR and QFD stages, while Groups B and C adapted more quickly, leveraging structured criteria to guide reasoning. This variation reflects a learning curve associated with matrix-based tools and unfamiliar evaluation logic.

Despite initial friction, participants reported increased confidence and fluency as the workshop progressed. Feedback highlighted the clarity offered by the structured criteria, especially in the MSASM and ADCD. However, participants suggested improvements such as simplified scoring logic, pivoted worksheet structures, and better integration with CAD or PLM tools. Preference for Pugh matrices over numeric scoring also indicated a need for more intuitive evaluation mechanisms.

Applicability was evident in how participants translated DFA and DFD principles into concrete modular design strategies. Group B prioritised assembly efficiency via simplified interfaces, tool reduction, and part orientation, while Group C emphasised accessibility, connector standardisation, and damage prevention. The resulting artefacts demonstrated alignment with their respective objectives, even though each group focused exclusively on one domain. This suggests the tool effectively framed early trade-offs and operational constraints.

The ADCD and MSASM enabled identification of critical pain points (e.g., M05 housing in Group A) and supported design iteration. Notably, although groups were focused on either assembly or disassembly, they frequently acknowledged the implications for both—indicating that the tool encouraged broader lifecycle thinking even in focused tasks.

### 6.2. Methodological considerations and limitations

The workshop gave triangulated data through behavioural (task time, field notes), cognitive (matrix use), and reflective (survey) insights. The multi-modal approach strengthens the credibility of findings. However, group segmentation limited evaluation of how the tool supports integrated design trade-offs between assembly and disassembly. Like other structured design methods, the tool introduces a cognitive load that must be managed. Introductory training and simplified templates could ease professional deployment.

The study was also constrained by a small sample size (n=6) and a single product case. Although participants were representatives of trained engineering students, broader validation across industrial settings and product categories is needed. This study is an early-stage validation aimed at exploring tool usability and applicability in a controlled academic setting. The sample size reflects a pilot-scale effort intended to guide further industrial testing.

Thematic analysis was performed by a single researcher, and future studies should include inter-rater coding to improve transparency and replicability.

### 6.3. Contributions and future work

This study gives a first structured evaluation of an expanded MFD that embeds production and end-of-life considerations into early modular product architecture. It contributes to design support tool literature by offering a structured framework for evaluating DFA/DFD implications during conceptual modularisation. The ADCD and MSASM support early visibility of design trade-offs typically delayed to later stages.

The aim of this study was not to assess the overall quality of the final modular concepts but to examine whether the expanded MFD tool effectively enabled early-stage reasoning aligned with DFA and DFD. A more comprehensive evaluation of design outcomes will be pursued in the future.

Further work should also involve larger samples and diverse industry cases to evaluate scalability and impact on outputs. Comparative studies between teams could reveal performance gaps. Integration with digital platforms (e.g., PALMA®) and CAD/PLM software could improve usability and adoption.

## 7. Conclusion

This paper introduced and evaluated an expanded MFD method that integrates DFA and DFD principles into early-stage modular product design. A workshop-based study demonstrated the tool's usability and applicability, highlighting its value for guiding structured reasoning about module architecture, interfaces, and production constraints.

Participants used visual and scoring tools to analyse trade-offs aligned with assigned design focus. While the tool had a learning curve, especially in early stages, it ultimately enabled deeper engagement with modularity concepts. Results suggest that structured, criteria-based evaluation can support early design reasoning in both academic and industrial contexts.

Future iterations will focus on improving onboarding, digital integration, and broader applicability to other design drivers such as cost, user experience, or maintainability. The approach is complementary to existing modularisation tools and can be integrated into broader design workflows where lifecycle performance is a priority.

## Acknowledgements

This study was part of a research collaboration between the main author at KTH and Sweden Modular Management AB. The author sincerely thanks Elia Martinelli and Jan Stamer for their support during the workshop, and Adam Lundström for contributing to participant recruitment.